# Environmental effects on the electrical characteristics of back-gated WSe$_2$ field effect transistors

**Francesca Urban[1,2], Lisanne Peters[3], Nadia Martucciello[2], Niall McEvoy[3] and Antonio Di Bartolomeo[1,2*]**

[1] Physics Department "E. R. Caianiello" and Interdepartmental Centre NanoMates, University of Salerno, via Giovanni Paolo II n. 132, Fisciano 84084, Italy

[2] CNR-SPIN Salerno, via Giovanni Paolo II n. 132, Fisciano 84084, Italy

[3] AMBER & School of Chemistry, Trinity College Dublin, Dublin 2, Ireland

*E-mail: adibartolomeo@unisa.it

**Abstract**

We study the effect of polymer coating, pressure and temperature on the electrical characteristics of monolayer WSe$_2$ back-gated transistors with quasi-ohmic Ni/Au contacts. We find that the removal of a layer of poly(methyl methacrylate) or decreasing the pressure change the device conductivity from p to n-type. We study the current-voltage characteristics as a function of the temperature and measure a gate-tunable Schottky barrier at the contacts with a height of ~60 meV in flat-band condition. We report and discuss a change in the mobility and the subthreshold slope observed with increasing temperature. Finally, we estimate the trap density at the WSe$_2$/SiO$_2$ interface and study the spectral photoresponse of the device, achieving a responsivity of ~$0.5\ AW^{-1}$ at $700\ nm$ wavelength and $0.37\ mWcm^{-2}$ optical power.

## 1. Introduction

The continuous downscaling of the channel length and thickness in modern field effect transistors (FETs) underlines the need for atomically layered materials to minimize short channel effects at extreme scaling limits [1,2]. Layered transition metal dichalcogenides (TMDs), owing to their two-dimensional structure, decent mobility and absence of dangling bonds, can enable extreme channel length scaling and have recently emerged as promising materials for future electronic and optoelectronic devices [3–7]. These graphene-like materials offer the advantages of sizeable and non-zero bandgap, high *on/off* ratio and quasi-ideal subthreshold swing, mechanical flexibility, and thermal and chemical stability. As for graphene, their electronic transport properties are strongly influenced by the choice of the metal contacts [8,9], interface traps and impurities [10,11], structural defects and environmental exposure [12–14]. These effects need to be understood and controlled for technological applications.

Molybdenum disulfide ($MoS_2$) has been one of the most heavily investigated systems from the TMD family to date [15–18]. Similar to $MoS_2$, tungsten diselenide ($WSe_2$), whose electrical and optical properties have been less explored [20], is characterized by an indirect bandgap (1.0-1.2 eV) in its bulk form and shows a transition to a direct gap of 1.6 eV when it is thinned to monolayer [21]. A few recent reports on $WSe_2$ FETs demonstrated relatively high field-effect mobility controllable by temperature and bias voltage [21], an ideal subthreshold slope $\sim 60\ mV/dec$ [22] and an *on/off* ratio greater than $10^8$. The ambipolar behavior, controllable using different metal contacts like In or Pd [8], which favour electron and hole injection respectively [23,24], makes mono- and few-layer $WSe_2$ an interesting material for complementary logic applications and a stable $WSe_2$- based CMOS technology has been proved [25].

As for other low-dimensional structures, a great challenge for electronic integration $WSe_2$ is the achievement of low-resistance ohmic contacts, a task often complicated by the appearance of Schottky barriers due to the occurrence of Fermi level pinning [26,27]. Accordingly, several studies have aimed at clarifying the role of the contacts, focusing on the carrier transport at the $WSe_2$/metal interface [8,25,26,28].

In this paper, we study back-gate monolayer $WSe_2$ devices with Ni contacts, measuring their electrical characteristics under different conditions, considering for instance the effect of a poly(methyl methacrylate) (PMMA) coating layer, and the dependence on the chamber pressure and the sample temperature. Similar to graphene [29,30], we observe that PMMA strongly influences the electric transport [31,32] to the extent of modifying the polarity of the device from p-type to n-type conduction when the PMMA layer is removed. We demonstrate that lowering the pressure on air exposed $WSe_2$-FETs affects their characteristics in a similar way to PMMA, turning the conduction from p- to n-type. Furthermore, from the current-voltage (I-V) characteristics measured at different temperatures, we are able to estimate the low Schottky barrier (SB) formed by the Ni contacts on the $WSe_2$ channel [33]. Then, we study the temperature dependence of the carrier mobility and the subthreshold slope [34], and show that both undergo a change of behavior with rising temperature. Furthermore, from the subthreshold slope data, we derive the interface trap density in the device. Finally, we measure the photo-response at several laser wavelengths, achieving a responsivity as high as $\sim 0.5\ AW^{-1}$ at 700 nm, i.e. at photon energy close to the $WSe_2$ bandgap.

## 2. Experimental

The $WSe_2$ flakes were grown by chemical vapor deposition (CVD) at 900°C and pressure of 6 Torr on highly p-doped Si (silicon) substrate covered by 300 nm of $SiO_2$ (silicon dioxide). Raman characterization with an unpolarized incident laser at of 532 $nm$ was performed to select monolayer flakes. An example of Raman spectrum is shown in Figure 1 (a) and exhibits two peaks around $\sim 250\ cm^{-1}$ and $\sim 260\ cm^{-1}$, corresponding to an overlapping contribution from the in-plane vibrations of W and Se atoms ($E^1_{2g}$) and out-of-plane vibration of Se atoms ($A_{1g}$), and to a second order resonant Raman mode ($2\ LA\ (M)$) due to LA phonons at the M point in the Brillouin zone [35,36], respectively. The peak frequency positions are typical of a $WSe_2$ monolayer of thickness $d \sim 0.7\ nm$ [24,36,37]. The monolayer structure of the flake is further verified by its response to laser light, as reported in the following.

The triangular shape of the $WSe_2$ monolayer channel and the evaporation-deposited Ni/Au ($5/50\ nm$) contacts are visible in the optical microscope image of Figure 1 (b), while Figure 1 (c) shows a schematic view of the back-gate FET device.

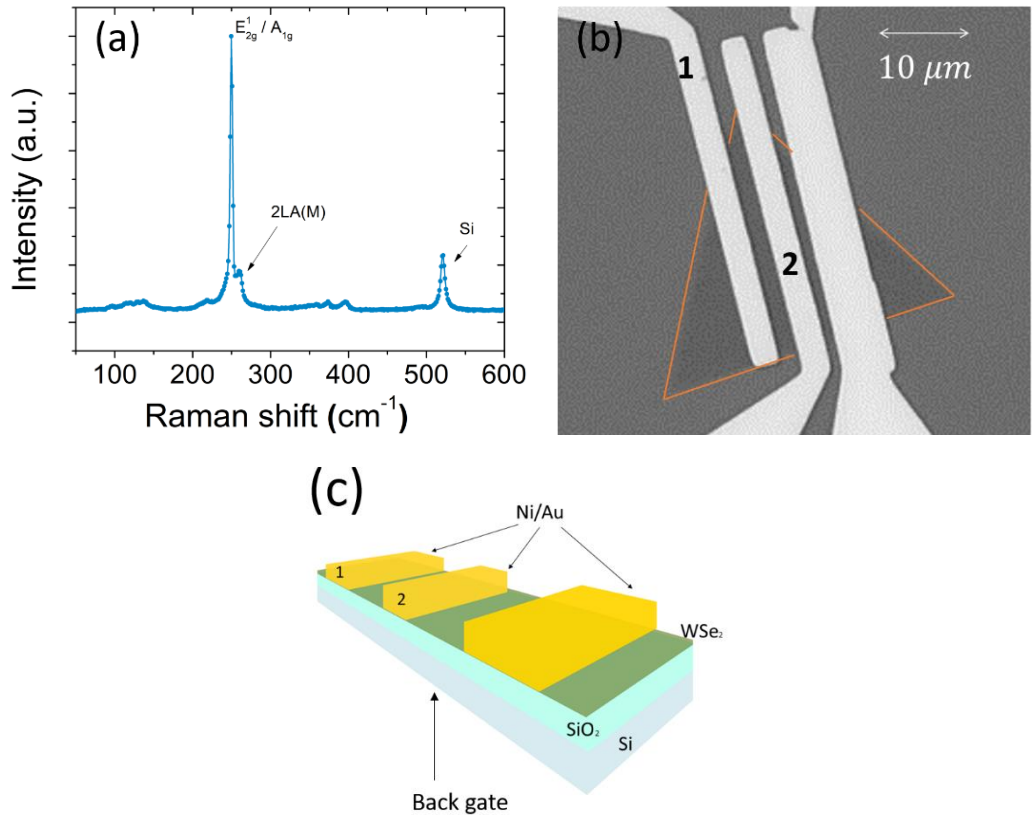

**Figure 1**(a) Raman spectrum of the WSe$_2$ flake used as the channel of the back gate FET transistor, whose optical microscope image and schematic section are shown in (b) and (c), respectively.

The electrical analyses are performed using a Keithley 4200 SCS (semiconductor characterization system) connected with a Janis ST-500 probe station, equipped with four probes used for the electrical connection to the drain and source Ni/Au terminals and to the Si back-gate of the device. In the following, the transistor characterization refers to contact 1 and 2, as marked in Figure 1 (b) and (c). The distance between the two contacts, i.e. the channel length, is L∼2 μm, while the mean channel width is W∼22 μm (Figure 1 (b)).

## 3. Results and discussion

We start presenting the transistor characterization by comparing the device electrical I-V curves with and without a PMMA coating layer. It has been observed that a PMMA film, or even only residues of it, cause p-type doping of graphene channels due to the presence of oxygen molecules which constitute the polymeric film. The p-type conduction is explained considering the charge transfer to $H_2O/O_2$ which, acting as electron capture centers, preserve the hole conduction but suppress the free electron density [27,39–41]. Here, we report a similar effect for WSe$_2$ FETs. We encapsulated the transistor channel with a layer of PMMA to protect it from residues and adsorbates and to potentially improve its electrical properties [31,32]. The PMMA-covered devices show p-type behavior, as can be seen from the $I_{ds} - V_{gs}$ transfer curves of Figure 2 (a) showing high channel current $I_{ds}$ (*on*-state of the FET) at negative gate voltages, $V_{gs}$. After the removal of the PMMA in $(CH_3)_2CO$ (acetone), a dramatic change to n-type behavior occurred, with *on*-state at $V_{gs} < 0\ V$, as shown in Figure 2 (b). We remark that a similar effect has been reported in [25], but for exfoliated WSe$_2$ flakes on an $SiO_2/Si$ substrate covered by $F_4PCNQ$ doped PMMA. The corresponding $I_{ds} - V_{ds}$ output characteristics are reported in Figure 2 (c) and (d); both of these show a slight asymmetric behavior, which is more evident when the gate voltage bias sets the device in the *off* state. This is an indication of low Schottky barriers at Ni/WSe$_2$ contacts [28,42].

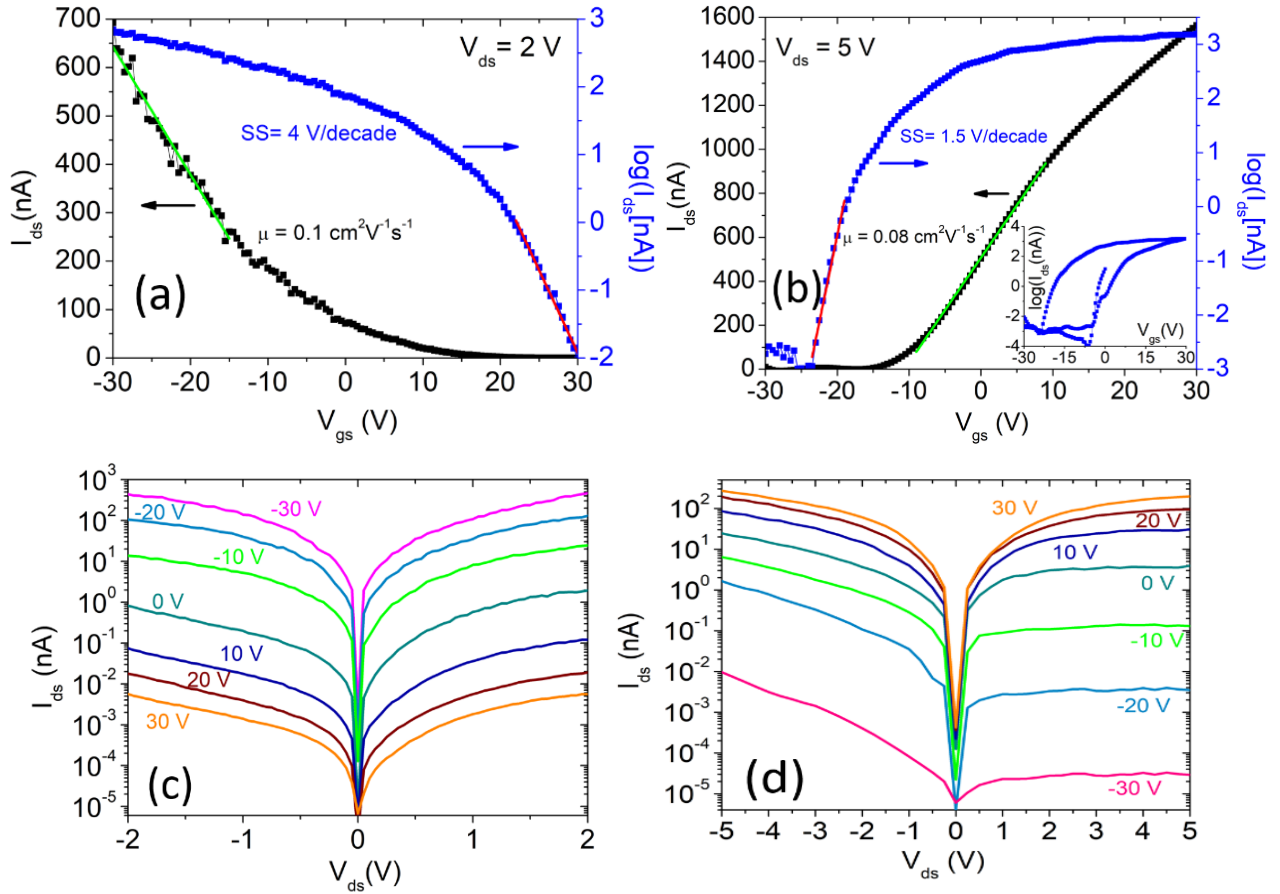

**Figure 2.** (a) Transfer characteristic ($I_{ds} - V_{gs}$ curves) performed at a drain voltage bias $V_{ds} = 2\,V$ for the device covered with PMMA. (b) Transfer characteristic performed at a drain voltage bias of 5 V after the removal of PMMA. The inset shows a complete cycle with the gate voltage $V_{gs}$ swept forward and backward. Output characteristics ($I_{ds} - V_{ds}$ curves) at different gate voltages for the device with (c) and without (d) PMMA. For the uncovered device, the drain bias was increased from $V_{ds} = 2V$ to $V_{ds} = +5V$ to better characterize the above-threshold region. All measurements were performed at $T = 293$ K and $P = 2.3$ mbar.

For increasing $V_{gs}$, the channel current at constant $V_{ds}$ bias shows an exponential dependence (below threshold region) followed by a linear or power law behavior (above threshold region).

A quadratic behavior is particularly evident in the transfer characteristic of Figure 2 (a), despite the transistor is operated in the triode region. The parabolic dependence of $I_{ds}$ on $V_{gs}$ can be ascribed to a linear gate-voltage dependence of the mobility $\mu$ [43,44], that defines the drain current as

$$I_{ds} = \frac{W C_{ox} \mu}{L}(V_{gs} - V_{th})V_{ds} \quad (1.1)$$

with

$$\mu = \mu_B(V_{gs} - V_{th}) \quad (1.2)$$

in which $C_{ox} \approx 11.6\,nFcm^{-2}$ is the SiO$_2$ capacitance per unit area, $\mu_B$ represents the mobility per unit gate voltage and $V_{th}$ is the gate threshold voltage. The $V_{gs}$-dependent mobility can be explained considering that the increasing carrier density becomes more effective in screening Coulomb

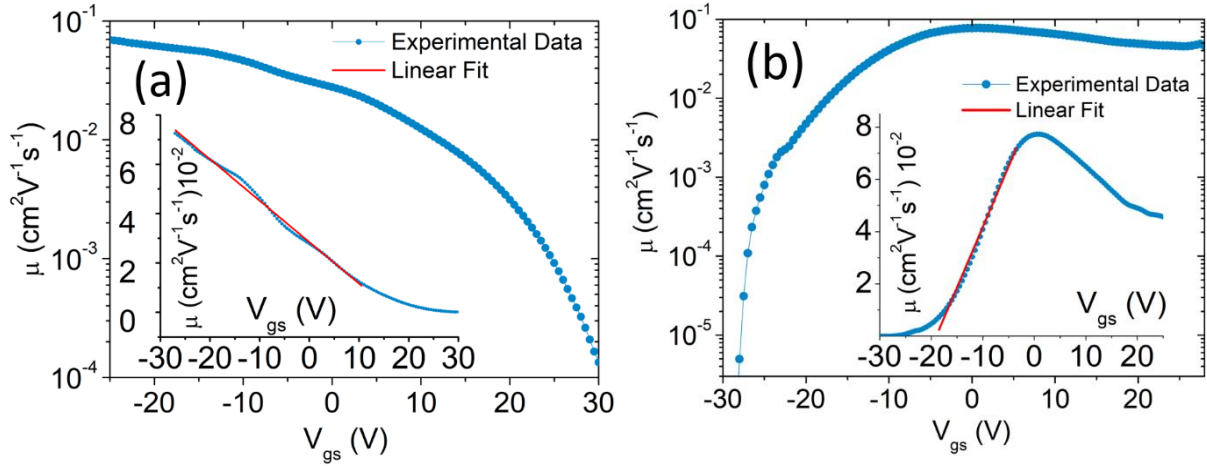

**Figure 3.** Mobility versus gate voltage on logarithmic scale for the $WSe_2$ flake covered (a) or uncovered (b) by PMMA. The inset graphs show the mobility on linear scale.

scatters or in filling trap states at higher $V_{gs}$, thus resulting in enhanced mobility.

According to eq. (1.1) and (1.2),

$$\mu(V_{gs}) = \frac{L}{W}\frac{1}{C_{ox}}\frac{1}{V_{ds}}\frac{dI_{ds}}{dV_{gs}} \quad (1.3)$$

Figure 3 (a) and (b) show the $\mu - V_{gs}$ curves on logarithmic and linear (insets) scales obtained from eq. (1.3) and the data of Figure 2, for the devices with and without PMMA, respectively. These confirm a linear dependence of $\mu$ on $V_{gs}$ over a certain range. Remarkably, for the device with removed PMMA, the mobility shows the typical decrease observed in common FETs due to increased scattering suffered by carriers attracted at the channel/dielectric interface at higher gate voltages.

Neglecting the $V_{gs}$ dependence of the mobility, as usually done in literature, $\mu$ can be obtained, by fitting a straight line to the transfer characteristic as shown in Figures 2 (a) and 2 (b). By this method, which confirms the previous results, we estimate an electron mobility of $\sim 0.08\ cm^2 V^{-1} s^{-1}$ for the n-type transistor without PMMA, consistent with other works with $WSe_2$ on $SiO_2$[41], and a hole mobility of $\sim 0.1\ cm^2 V^{-1} s^{-1}$ for the p-type transistor with covering PMMA, thus confirming the higher hole mobility in $WSe_2$ reported elsewhere [24,27,32,42].

The subthreshold slope $SS = dV_{gs}/dlog(I_{ds})$ is $4\ V/decade$ and $1.5\ V/decade$, for the p-type and n-type transistor, respectively. The different $SS$ results from a different trap density at the $WSe_2$/dielectric interface [45,46], implying a higher trap density when the $WSe_2$ channel is covered by PMMA, which adds a second interface. The trap states manifest also as a hysteresis in the transfer curve, as shown in the inset of Figure 2 (b), due to trapping and detrapping of charge carriers, whose potential adds to that of the back-gate [43,44,47,48].

After the removal of the polymeric film and exposure of the device to air for a few days, we observed a restoration of prevailing p-type behavior due to $O_2$ and water adsorption on the $WSe_2$ surface. Then, we studied the effect of dynamic pressure by increasing the vacuum level of the probe station chamber from the atmospheric value to $P = 10^{-5}$ mbar. As reported in Figure 4 (a), the transistor transfer characteristic changed again from p to n-type. The transition occurred with a gradual decrease of the subthreshold slope and an increase of the *on/off* ratio, as shown in Figure 4 (b). The polarity change is caused by desorption of adsorbed $O_2$ and $H_2O$, as suggested also in [30] where a similar effect was reported for $WS_2$ FETs.

The study of the electrical characteristics of the transistor as a function of temperature $T$ is useful to get information on the Schottky barrier height at $WSe_2$/Ni contacts.

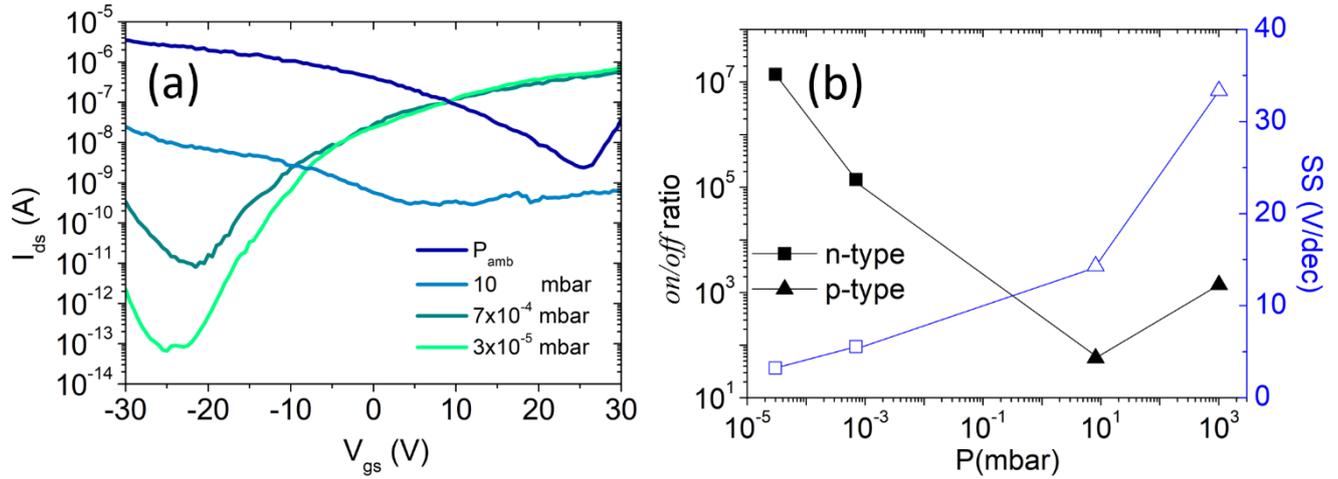

**Figure 4.** (a) Transfer characteristics at different pressures from atmospheric value (blue curve) to $\sim 10^{-5}$ mbar (light green curve) (b) Threshold voltage and subthreshold slope as a function of chamber pressure.

Following the procedure proposed by Das and Appenzeller [23,49,50], we extracted the Schottky barrier at the flat-band condition (Figure 5) from a plot of the Schottky barrier height evaluated as a function of $V_{gs}$. We started measuring the $I_{ds} - V_{gs}$ characteristics of the device at several temperatures as reported in Figure 5 (a). For given values of the gate voltage (two examples are marked by the vertical lines in Figure 5 (a)), we extracted $I_{ds} - T$ datasets that we used to construct the Arrhenius plot of Figure 5 (b). Such plot shows the $ln(I_{ds}) - \frac{1}{T}$ curves at a representative subset of $V_{gs}$ values. We assume that the contacts behave as Schottky junctions with a current-temperature dependence such that

$$I_{ds} \sim T^2 \, exp\left(-\frac{q\Phi_B}{kT}\right) \quad (1.4)$$

where $k$ is the Boltzmann constant, $q$ the electron charge, $T$ the absolute temperature and $\Phi_B$ the Schottky barrier height. By neglecting the $T^2$ (or $T^{\frac{3}{2}}$) factor with respect to the exponential [49,51], according to eq. (1.4), a linear fit of each $V_{gs}$ dataset in Figure 5 (b) yields a Schottky barrier $\Phi_B$. The so-obtained $\Phi_B - V_{gs}$ relationship is displayed in Figure 5 (c) and can be divided into three zones, each one corresponding to a different transport regime, consistent with the behavior of the transfer characteristics of Figure 5 (a).

At low gate voltage, the device is set in the *off* state and the transport is due to the thermal excitation of electrons over the barrier, which is gradually lowered by the gate, as sketched in the insets of Figure 5 (c). This corresponds to a steep exponential rise of the current in the transfer characteristics (with $60 \frac{mV}{decade}$ slope in the ideal case). When the gate voltage is further increased, the device reaches the flat band condition ($V_{gs} = V_{FB}$), which sometimes appears in the subthreshold part of the transfer characteristics as a sudden change of slope; for $V_{gs} > V_{FB}$ the device enters the so-called Schottky regime which includes part of the downward bended region of $I_{ds} - V_{ds}$ curves and is characterized by thermionic and thermionic field emission; finally, at higher $V_{gs}$, tunneling through the thinned Ni/WSe$_2$ barrier becomes the dominant conduction mechanism and the device reaches the region above-threshold characterized by a $I_{ds} - V_{gs}$ power-law dependence.

The gate voltage, that corresponds to $V_{FB}$, is identified by the change of slope in the $\Phi_B - V_{gs}$ plot at lower $V_{gs}$ (marked by the "Flat band" vertical line of Figure 5 (c)). The $\Phi_B$ corresponding to $V_{FB}$ is the so-called flat-band Schottky barrier and is $\sim 60 \, meV$, pointing to quasi-ohmic Ni-WSe$_2$ contacts, consistent with the low rectifying output characteristics shown in Figure 2.

Figure 5 (d) shows the temperature behavior of the threshold voltage $V_{th}$, which has been extracted assuming a quadratic $I_{ds} - V_{gs}$ law as expressed by eq. (1.1) and (1.2). The decreasing behavior is easily explained considering that the rising temperature accelerates the transition from the Schottky to the power-law regime; the plot seems to indicate a change of slope around room temperature.

Figure 6 (a) shows the temperature behavior the mobility $\mu$ at $V_{gs} = 10 \, V$ obtained from the same quadratic fits of the $I_{ds} - V_{gs}$ curves.

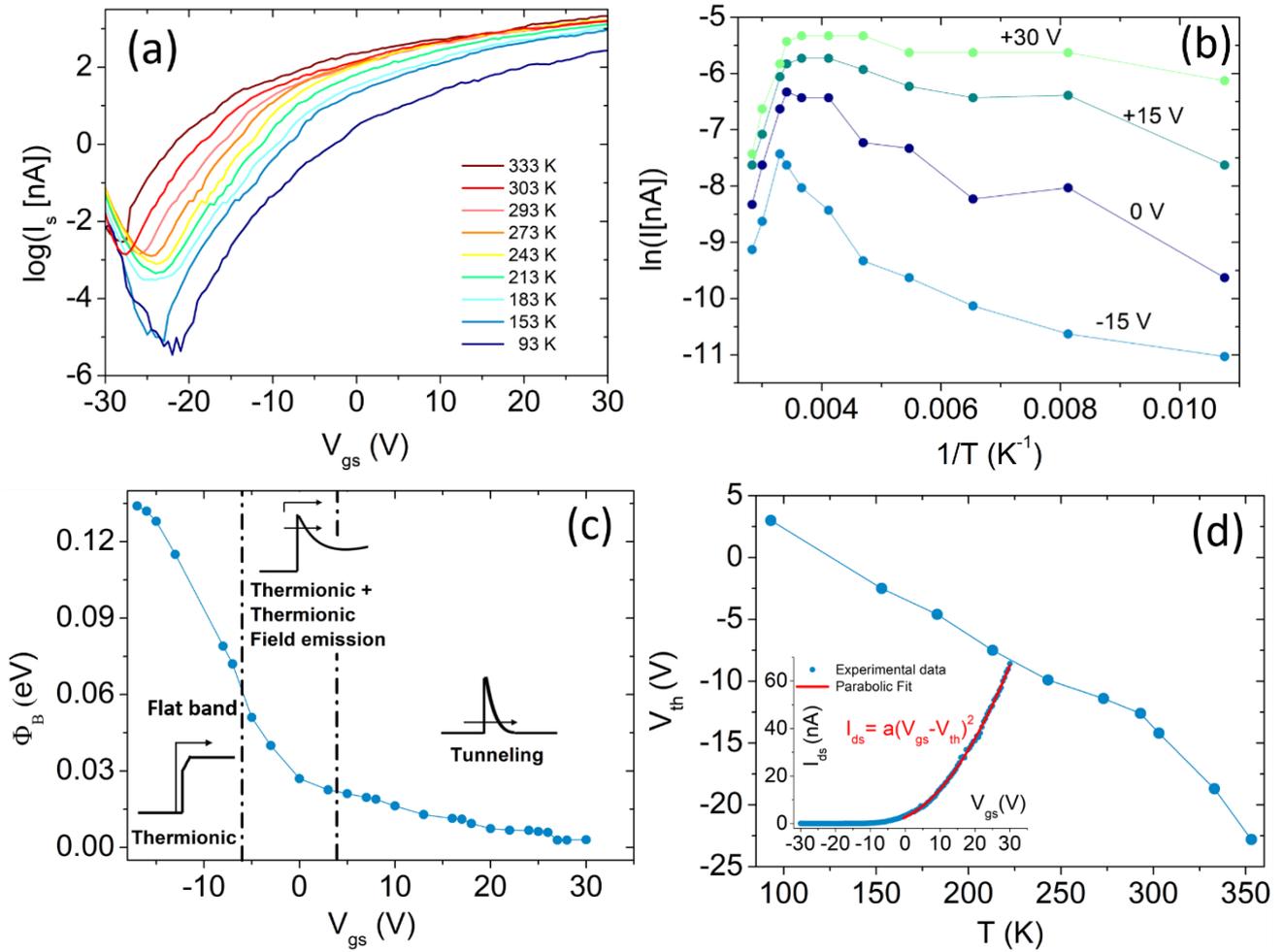

**Figure 5.** (a) Transfer characteristic at different temperatures. (b) Arrhenius plot of the current at different temperatures corresponding to a subset of the gate voltages (two of these $V_{gs}$ values are represented by the vertical lines in (a)). (c) Apparent Schottky barrier as a function of the gate voltage; the insets show the band alignment and the transport regimes at the Ni/WSe$_2$ contacts. (d) Threshold voltage $V_{th}$ as a function of the temperature; the inset shows, as an example, the parabolic fit of the $I_{ds} - V_{gs}$ curve at $T = 273\,K$.

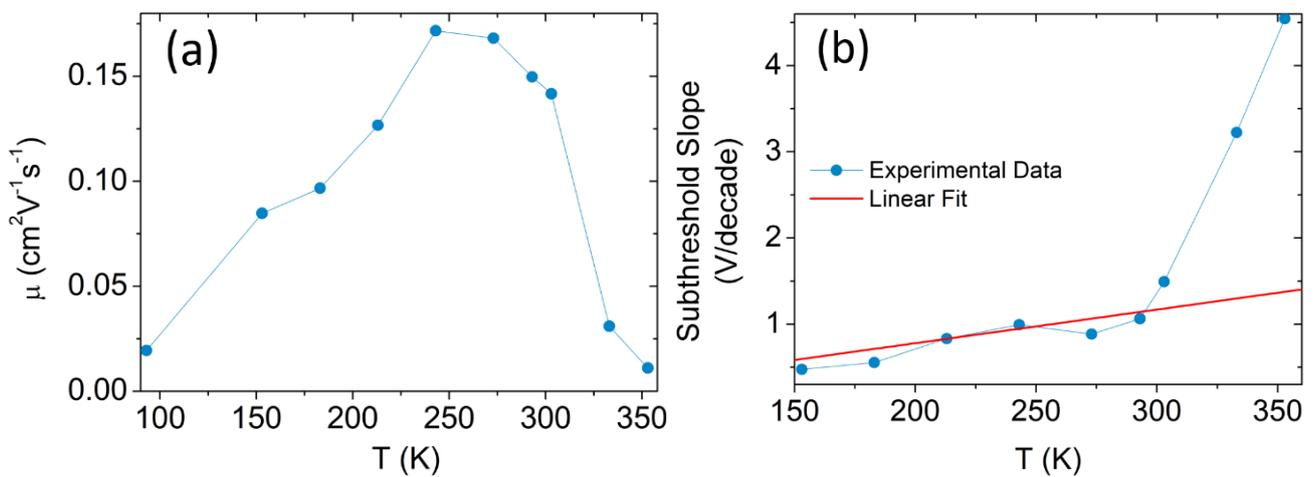

**Figure 6.** Temperature dependence of (a) mobility $\mu$ and (b) subthreshold slope *SS*.

The mobility shows a crossover at $T \sim 250\,K$ between a positive and a negative $\frac{d\mu}{dT}$, which is a typical behavior of semiconductor materials. This corresponds to a change from charged-impurity coulomb scattering, dominating at lower temperatures, to phonon scattering becoming the conduction limiting mechanism at higher temperature [31].

The subthreshold slope has a dependence on the temperature that can be simplified with the following expression

$$SS = n \frac{kT}{q} ln10 \qquad (1.5)$$

where $n$ is the body factor which is related to the interface trap ($C_{it}$) and the channel depletion-layer ($C_{dl}$) capacitances by

$$n = 1 + \frac{C_{it} + C_{dl}}{C_{ox}} \qquad (1.6)$$

Figure 6 (b) shows the expected linear $SS - T$ dependence with a sudden increase above room temperature. The deviation from the expected behavior is an effect of the low Schottky barrier which becomes ineffective above room temperature ($kT = 26\,meV$), resulting in an increase of the subthreshold current leakage.

Assuming that the WSe$_2$ monolayer channel is fully depleted, i.e. $C_{dl} \approx 0$, from the fit of the experimental data with equation (1.5), we obtain a $n \approx 48$ and an interface trap density $N_{it} = \frac{C_{it}}{q^2} \approx 1.3 \times 10^{13}\,eV^{-1}cm^{-2}$, which is consistent with previous results reported in the literature [52].

The presence of such a density of trap states justifies the observed hysteretic behavior of the transfer characteristic, inset of Figure 2 (b) [44].

As a further characterization of the device, we performed photocurrent measurements with light at different wavelengths, selected by filtering a supercontinuous laser source (NKT Photonics, Superk Compact, wavelength ranging from 450 nm to 2400 nm, total output power of $110\,mW$) using pass-band filters with 50 nm bandwidth. Figure 7 shows the photoresponse of the WSe$_2$ FET for five different laser wavelengths. The photocurrent exhibits a higher peak at the wavelength of 700 nm (photon energy 1.7 eV), which is slightly above the bandgap of a WSe$_2$ monolayer, confirming the Raman spectroscopy assignment of the single-layer nature of the WSe$_2$ channel.

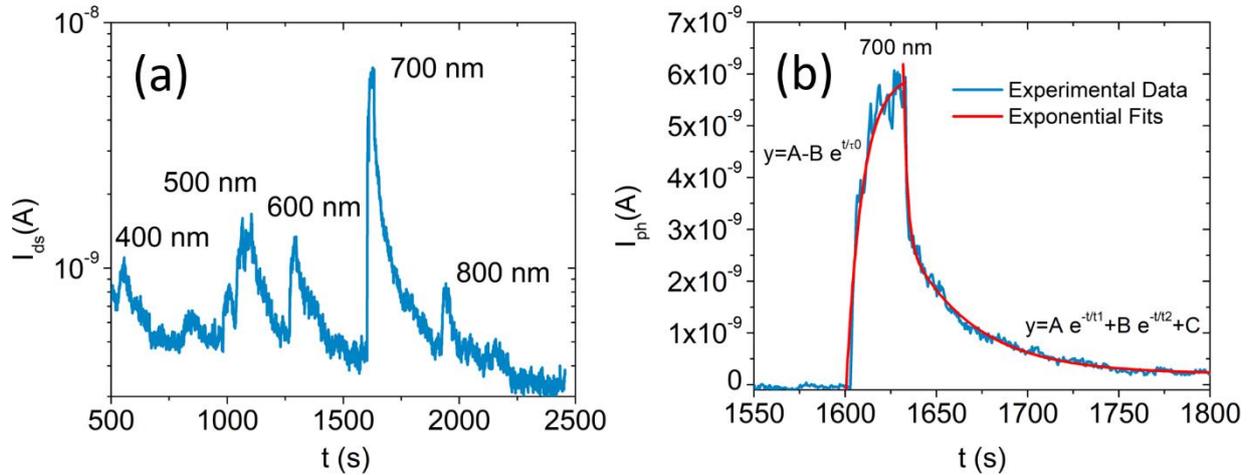

**Figure 7.** (a) Drain-Source current measured under illumination at different wavelengths ($V_{ds} = -5\,V$, $V_{gs} = 0\,V$, $P \sim 10^{-4}\,mbar$). (b) Photocurrent generated by a $30\,s$ laser pulse at the wavelength of $\sim 700\,nm$ and optical power $\sim 0.37\,mW/cm^2$ with exponential fits.

Figure 7 (b) reports the photocurrent, $I_{ph} = I_{light} - I_{dark}$, obtained in response to a laser pulse of $30\,s$ at the wavelength of $\sim 700\,nm$, and optical power about $0.37\,mWcm^{-2}$. It corresponds to a peak with rising time $\tau_0 \sim 9\,s$ and a double exponential decay with times $\tau_1 \sim 2\,s$ and $\tau_2 \sim 36\,s$. The double decay indicate the presence of faster and slow traps [53] and is consistent with a photoresponse speed longer than 5 s for quasi-ohmic contacts on similar WSe$_2$ FETs [50]. We notice that the contact type can play an important role in the response time of WSe$_2$ phototransistors; indeed reduced times have been reported for Schottky contacts [54,55].

From the data in Figure 7 (b), we estimate a photoresponsivity

$$R = \frac{I_{ph}}{W_{opt}} \approx 0.5 \frac{A}{W} \qquad (1.7)$$

where $W_{opt}$ is the incident power. This value is in good agreement with the 0.6 $A/W$ at 750 nm previously reported [55]. Such a responsivity is competitive with solid-state devices on the market, despite the ultrathin absorber. It confirms the excellent photoresponse of monolayer WSe$_2$ enhanced by the direct bandgap [56,57]

## 4. Conclusions

In conclusion, we showed that different environmental conditions can have dramatic effects on the electrical properties of monolayer WSe$_2$ back-gated transistors. In particular, we demonstrated that the removal of a polymer coating layer, as well as of oxygen and water adsorbates, can change the conduction from p to n-type. From I-V characterization at different temperatures, we extracted the Ni/WSe$_2$ Schottky barrier height, which we studied as a function of the back-gate voltage. We reported and discussed a change in the temperature behavior of the mobility and the subthreshold slope. Finally, we studied the photoresponse of the device to selected laser wavelengths showing a prominent peak of photocurrent corresponding to the WSe$_2$ monolayer bandgap.

**Acknowledgments**

We acknowledge the economic support by POR Campania FSE 2014–2020, Asse III Ob. specifico l4, D.D. n. 80 del 31/05/2016 and CNR-SPIN SEED Project 2017.